\documentclass[iop]{emulateapj}
\usepackage{amsmath}
\usepackage{apjfonts}
\usepackage{subfigure}
\usepackage{booktabs}
\usepackage{color}
\usepackage{graphicx}
\usepackage{epstopdf}
\usepackage[toc,page,title,titletoc,header]{appendix}
\usepackage{ulem}
\usepackage{amsmath}

\begin{document}



\title{chemical abundances and ages of the bulge stars in APOGEE high-velocity peaks}

\author{Yingying Zhou\altaffilmark{1,2},  Juntai Shen\altaffilmark{*1,2},  Chao Liu\altaffilmark{3}, Zhao-Yu Li\altaffilmark{1,2}, Shude Mao\altaffilmark{4,3,5}, Andrea Kunder\altaffilmark{6}, R. Michael Rich\altaffilmark{7}, G. Zasowski\altaffilmark{8,9}, J. G. Fern\'{a}ndez-Trincado\altaffilmark{10}, Steven R. Majewski\altaffilmark{11}, Chien-Cheng Lin\altaffilmark{1,12},   Doug Geisler\altaffilmark{10}, Baitian Tang\altaffilmark{10}, S. Villanova\altaffilmark{10}, A. Roman-Lopes\altaffilmark{13}, M.  Schultheis\altaffilmark{14},  David L. Nidever\altaffilmark{15},  Andr\'{e}s Meza\altaffilmark{16}, Kaike Pan\altaffilmark{17},D. V. Bizyaev\altaffilmark{17,18}}

\affil{$^1$Key Laboratory for Research in Galaxies and Cosmology, Shanghai Astronomical Observatory, Chinese Academy of Sciences, 80 Nandan Road, Shanghai 200030, China \\
$^2$College of Astronomy and Space Sciences, University of Chinese Academy of Sciences, 19A Yuquan Road, Beijing 100049, China\\
$^3$Kay Lab of Optical Astronomy, National Astronomical Observatories, Chinese Academy of Sciences, Beijing 100012, China \\
$^4$Center for Astrophysics , Department of Physics, Tsinghua University, 10086, Beijing, China\\
$^5$Jodrell Bank Centre for Astrophysics, School of Physics and Astronomy, The University of Manchester, Oxford Road, Manchester M13 9PL, UK \\
$^6$ Leibniz-Institut f\"{u}r Astrophysik Potsdam (AIP), An der Sternwarte 16, D-14482 Potsdam, Germany \\
$^7$ Department of Physics and Astronomy, University of California at Los Angeles, Los Angeles, CA 90095-1562, USA \\
$^8$ Space Telescope Science Institute 3700 San Martin Drive, Baltimore, MD 21218, USA\\
$^9$ University of Utah, Department of Physics and Astronomy, 115 S. 1400 E., Salt Lake City, UT 84112, USA \\
$^{10}$ Departamento de Astronom\'{i}a, Casilla 160-C, Universidad de Concepci\'{o}n, Concepci\'{o}n, Chile \\
$^{11}$ Department of Astronomy, University of Virginia, P.O. Box 400325, Charlottesville, VA 22904-4325, USA\\
$^{12}$ Max Planck Institute for Astronomy, K{\"o}nigstuhl 17 17, D-69117 Heidelberg, Germany\\
$^{13}$ Departamento de F\'{i}sica, Facultad de Ciencias, Universidad de La Serena, Cisternas 1200, La Serena, Chile \\
$^{14}$ Laboratoire Lagrange, Universit\'{e} C\^{o}te d'Azur, Observatoire de la C\^{o}te d'Azur, CNRS, Blvd de l'Observatoire, F-06304 Nice, France \\
$^{15}$ National Optical Astronomy Observatory, 950 North Cherry Avenue, Tucson, AZ 85719, USA\\
$^{16}$Facultad de Ingenie\'{i}a, Universidad Aut\'{o}noma de Chile, Pedro de Valdivia 425, Santiago, Chile\\
$^{17}$Apache Point Observatory and New Mexico State University, P.O. Box 59, Sunspot, NM, 88349-0059, USA\\
$^{18}$Sternberg Astronomical Institute, Moscow State University, Moscow, Russia\\
$^{*}$Correspondence should be addressed to: jshen@shao.ac.cn }

\begin{abstract}

A cold, high-velocity (HV, $\sim$ 200 $\rm km\,s^{-1}$) peak  was first reported in several Galactic bulge fields based on the Apache Point Observatory Galaxy Evolution Experiment (APOGEE) commissioning observations. Both the existence and the nature of the HV peak are still under debate. Here we revisit this feature with the latest APOGEE DR13 data.  We find that most of the low latitude bulge fields display a skewed Gaussian distribution with an HV shoulder. However, only 3 out of 53 fields show distinct HV peaks around 200 $\rm km\,s^{-1}$.
The velocity distribution can be well described by Gauss$-$Hermite polynomials, except for the three fields showing clear HV peaks. We find that the correlation between the skewness parameter ($h_{3}$) and the mean velocity  ($\bar{v}$), instead of a distinctive HV peak, is a strong indicator of the bar. It was recently suggested that the HV peak is composed of preferentially young stars. We choose three fields showing clear HV peaks to test this hypothesis using the metallicity, [$\alpha$/M] and [C/N] as age proxies. We find that both young and old stars show HV features. The similarity between the chemical abundances of stars in the HV peaks and the main component indicates that they are not systematically different in terms of chemical abundance or age. In contrast, there are clear differences in chemical space between stars in the Sagittarius dwarf and the bulge stars. The strong HV peaks off-plane are still to be explained properly and could be different in nature.
\end{abstract}

\section{introduction}
Bulges, disks, and halos are the main components in most spiral galaxies. Bulges are very common; they can be found in more than 80\% of Milky Way (MW) size galaxies \citep{FisherandDrory2011}. There are two main types of bulges: pseudo-bulges (disk-like, rotation dominated) and classical bulges (mini-elliptical, random motion dominated).  
The near infrared images from the \textit{COBE} satellite reveal clearly that the MW contains an asymmetric parallelogram-shaped boxy bulge in the center \citep{Weiland1994}. The stellar mass of the MW bulge in the volume of the red clump giant (RCG) measurement ($\pm2.2 \times \pm 1.4 \times \pm1.2$ kpc) is $\sim 1.25-1.6 \times 10^{10} M_{\odot}$ and the bulge composes $\sim$ 20\% of the total Galactic luminosity \citep{Portail2015}. Using red clump giants from the OGLE survey in the bulge region ($-10 < l < 10$, $2 < |b| < 7$), \cite{Cao2013} constructed a Galactic bulge model with total bar mass of $\sim$ 1.8 $\times 10^{10}\rm M_{\odot}$. Increasingly more evidence supports that the bulk of our MW bulge is probably a boxy/peanut-shaped bar (see, e.g., the recent review by \citealt{Shenli2016}).

A cold ($\sigma_{V} \sim 30$ $\rm km\,s^{-1}$) high-velocity (HV; $V_{\rm GSR} \approx +220$ $\rm km\,s^{-1}$) peak in the bulge radial velocity distribution was first reported by \cite{Nidever2012} with the Apache Point Observatory Galaxy Evolution Experiment (APOGEE; \citealt{Majewski2017}) commissioning data (first released in DR10). 
Based on the Giraffe Inner Bulge Survey (GIBS), \cite{Zoccali2014} and \cite{Zoccali2016} found no significant HV peaks in the bulge. However, the APOGEE HV detection was predominantly in the lowest-latitude fields and GIBS did not observe in-plane fields ($b=0^{\circ}$). The possible  HV peaks have raised intense discussions about its nature, which is still unclear. 
One potential explanation is that these stars are part of the tidal tail of Sagittarius (Sgr), which lies near the bulge fields, but the distances of the HV stars are roughly from 5 to 10 kpc, which makes this explanation unlikely \citep{Nidever2012}. The HV peaks are not likely to be a new substructure in the halo. \cite{Nidever2012} suggested that the HV stars are most likely bulge stars on bar orbits.
Analysis of the bulge orbits showed that the HV peak can be due to the motion of stars in the bar-supporting 2:1 resonant family and in other higher-order resonances \citep{Aumer2015, Molloy2015}. However, stars in the bar may not be on exactly resonant orbits.
Using the model in \cite{Shen2010}, \cite{Lizhaoyu2014} suggested that the full velocity distribution of the stars making up the bar potential can only produce an HV shoulder instead of a distinct HV peak. \cite{Lizhaoyu2014} speculated that the observed cold HV peak might be an artificial small number statistic. \cite{Debattista2015} suggested that the HV peak may be explained by a kiloparsec-scale nuclear stellar disk in the Galactic bulge. 
They predicted second peaks in the line-of-sight velocity distribution (LOSVD) at Galactic longitude $l=8^{\circ} \pm 2^{\circ}$, but not off the midplane.

\cite{Aumer2015} who argued that the selection function of APOGEE is more sensitive to young stars, found that the velocity distribution of young populations represents such an HV peak in their model. The HV peak could be due to young stars formed from gas just outside the growing bar and subsequently captured by it.

Stellar ages are important in understanding the HV peak, but they are hard to measure directly. 
Astroseismology provides a good approximation of age by measuring the time spent in the core hydrogen burning phase, which is a function of stellar mass, but it is observationally expensive and  currently available only for relatively small samples of stars, and it is not feasible to obtain for a large survey. 
However, stars with different ages can display some different chemical abundances. For example, the C and N elemental abundances in red giants are sensitive to stellar characteristics, including mass and age \citep{Masseron2015, Martig2015}.
APOGEE provides chemical information (including C and N) for stars with an internal precision of 0.05-0.09 dex, and with an external accuracy of 0.1-0.2 dex \citep{Holtzman2015}. Therefore, the elemental abundances derived from APOGEE giant stars can be used to infer ages. Chemical abundances and the inferred ages may provide important clues to the nature of the HV peaks/shoulders.

Our motivations of this work are to (1) revisit the HV peak in the latest APOGEE data release DR13, (2) see whether or not the stars in the HV peak have a different age composition compared to the main component, and (3) determine the age composition of the APOGEE bulge stars and whether the HV peak is more pronounced in young stars.

The paper is organized as follows. Section 2 introduces the selection method of our main sample. In Section 3 we check the velocity distribution of every field for the possible existence of the HV peak ($\sim$200 $\rm km\,s^{-1}$), fit the velocity distribution with Gauss$-$Hermite polynomials and compare the results to \cite{Shen2010} model. To study chemical abundance differences between the HV peak stars and main sample, stars are divided into two components depending on their velocities. In Section 4 we analyze the chemical abundance including [M/H], [$\alpha$/M] and [C/N] of the two components to see whether the HV stars have distinct chemical features and study the velocity distributions of the young and the old populations, using [C/N] as an age proxy. In Section 5 we summarize our main results.

\section{sample selection} 
Because of operating in the near-IR (1.51-1.70$\mu$m), APOGEE provides a good window to understanding the heavily dust-obscured bulge. APOGEE collects high-resolution (\textit{R} $\sim$ 22,500) spectra with a multiplexing (300 fibered objects) capability and provides a catalog with radial velocity, stellar parameters, and up to 15 elemental abundances for over 150,000 stars in DR12 \citep{Gunn2006, Holtzman2015, Majewski2017}. 

DR13 does not have any additional targets compared to DR12, but the data reduction and analysis have been improved in several ways. In this work we consider the bulge region covering the Galactic longitude $l$ from $-5^{\circ}$ to $20^{\circ}$ and Galactic latitude $b$ from $-20^{\circ}$  to $20^{\circ}$. The Sgr dwarf spheroidal (dSph) galaxy is also located in the region.  In this special field, Sgr member candidates are targeted based on a selection of Two Micron All Sky Survey (2MASS) M giants \citep{Majewski2003}.

There are $\sim$20,000 stars in the selected region. However, $\sim$10,000 stars do not have accurate effective temperature,  due to either the low signal-to-noise ratio $\rm S/N$ or the temperature limitations in the ASPCAP  pipeline \citep{Garc2016}.  The STARFLAG indicator is used to avoid bad pixels, and we do not consider stars with very bright neighbors or with low $\rm S/N$ ($\rm S/N<5$). To avoid globular cluster, binary and variable stars, we only consider velocity dispersions ($vscatter$) less than 1 $\rm km\,s^{-1}$ over four visits \citep{Nidever2015, Fern2016}. To exclude foreground dwarfs, we apply a
 $\rm log$$\;g$ < 3.8 limit \citep{Ness2016}.

APOGEE provides accurate line-of-sight velocities. 
The galactocentric standard of rest velocity ($V_{\rm GSR}$, the line-of-sight velocity that would be observed by a stationary observer at the Sun's position) is derived as follows: 
\begin{equation}
	V_{\rm GSR}=V_{\rm HC}+U_{\odot}{\rm cos}\,b\;{\rm cos}\,l+(V_{\odot}+239){\rm cos}\,b \;{\rm sin} \,l +W_{\odot}{\rm sin} \,b,
\end{equation}	
where $V_{\rm HC}$ is the heliocentric velocity provided in DR13. 
We adopt the solar peculiar motion ($U_{\odot}$, $V_{\odot}$, $W_{\odot}$) as (11.1$^{+0.69}_{-0.75}$, 12.24$^{+0.47}_{-0.47}$, 7.25$^{+0.37}_{-0.36}$) $\rm km\,s^{-1}$ \citep{Sch10}, and the circular speed of the local standard of rest as 239 $\pm$ 5 $\rm km\,s^{-1}$ \citep{McMillan2011}. We tested the other values of the solar motion, such as those of \cite{Tian2015} and \cite{Huangyang2015}, as well as the circular speed of 220 $\rm km\,s^{-1}$, and found that these different values would not change our conclusions.\footnote{The difference in ($U_{\odot}$, $V_{\odot}$, $W_{\odot}$) is smaller than 2 $\rm km\,s^{-1}$, which is quite small compared to our bin size (20 $\rm km\,s^{-1}$). Different rotation velocities at the solar radius will not change the shape of velocity distribution in each field because the field size is very small (2 $\rm deg^{2}$). It will only shift the whole profile systematically without changing its shape.}

\begin{figure*}[!t]
\centering
\includegraphics[width=1.8\columnwidth]{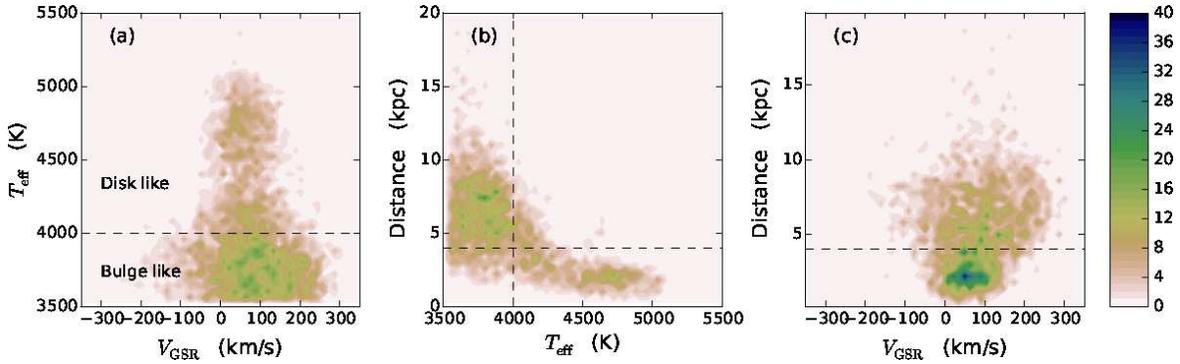}
\caption{(a) The $T_{\rm eff}$ - $V_{\rm GSR}$ distribution for the selected APOGEE bulge sample in the region $-5^{\circ} < l < 20^{\circ}, -3^{\circ} < b < 3^{\circ}$ and with distance measured.
For stars with high effective temperature ($T_{\rm eff}$ > 4000 K), the velocity dispersion is small (consistent with the disk-like kinematics), while lower-temperature stars have a larger velocity dispersion similar to bulge-like kinematics. The black dashed line shows our temperature cut to separate possible foreground disk stars from bulge stars. (b) Relation between distance and $T_{\rm eff}$ for the same stars. The vertical dashed line marks $T_{\rm eff}=$4000 K, and the horizontal dashed line corresponds to a distance of 4 kpc. A clear anticorrelation can be seen between distance and $T_{\rm eff}$. (c) The  $V_{\rm GSR}$ $-$ distance distribution for the sample. The dashed line corresponds to 4 kpc.}
\label{fig-vg-teff}
\vspace{0.2cm}
\end{figure*}

The $T_{\rm eff}$ - $V_{\rm GSR}$ distribution of stars located in bulge direction is shown in the left panel of Figure \ref{fig-vg-teff}. The $T_{\rm eff}$ distribution of low-velocity stars ($-$50 $\rm km\,s^{-1}$ < $V_{\rm GSR} < 150 $ $\rm km\,s^{-1}$) spreads over a wider temperature range (from 3500 K to 5000 K), while the $T_{\rm eff}$ of HV stars ($V_{\rm GSR}$ > 150 $\rm km\,s^{-1}$) spreads over a narrower range (from 3600 K to 4300 K). 
Due to our cut in $\rm log$$\;g$, most of the stars are giants, so the higher-$T_{\rm eff}$ stars are located generally on the lower part of the red giant branch (RGB) and thus have relatively lower luminosity, which leads to a selection effect that stars with higher $T_{\rm eff}$ are likely closer to the Sun than those with lower $T_{\rm eff}$. 
The smaller velocity dispersion for the stars with higher $T_{\rm eff}$ shown in Figure \ref{fig-vg-teff} is another indication that a larger fraction of those stars may be foreground disk stars. The much larger velocity dispersion for the cool stars ($T_{\rm eff}$ < 4000 K) also hints that the majority of this sample belongs to the Galactic bulge.

Distances are given in a value-added catalog of DR13 by  \cite{Santiago2016}. The middle and right panels in Figure \ref{fig-vg-teff} show the stellar distributions in distance$-T_{\rm eff}$ and distance$-V_{\rm GSR}$, respectively. The right panel is roughly consistent with the predicted distance$-$velocity diagram shown in \cite{Lizhaoyu2014} based on the \cite{Shen2010} model. This confirms our assertion that the lower-dispersion/higher $T_{\rm eff}$ giants are significantly more contaminated by foreground disk stars. At the temperature limit of 4000 K, the stellar distance ranges from about 4 to 12 kpc, with most of the foreground stars removed. Unfortunately, a significant fraction of the APOGEE stars do not have distances measured by \cite{Santiago2016}, so in the end we do not use the distance cut in this study. Instead, we use the temperature cut of 4000 K to exclude the foreground disk contamination. Note that the exact value of the temperature cut does not affect our conclusions. The temperature cut at 4000 K corresponds to $\rm log$$\;g$ cut of 1.5 since $\rm log$$\;g$ is roughly correlated with $T_{\rm eff} $ for the red giant branch.
There are 2065 stars with $T_{\rm eff} < $ 4000 K in Figure \ref{fig-vg-teff}; 319 of these are disk stars with distance less than 4 kpc. Thus, in this subsample, the fraction of disk contamination is 15$\%$. The observed and derived quantities, especially distances, have uncertainties, and thus the actual fraction of foreground (and also possibly background) disk stars in the cleaned sample might be higher than the estimated 15$\%$.

\section{kinematical results}
\subsection{Velocity distributions in the Galactic bulge} 
\begin{figure*}[!t]
\centering
\includegraphics[width=1.8\columnwidth]{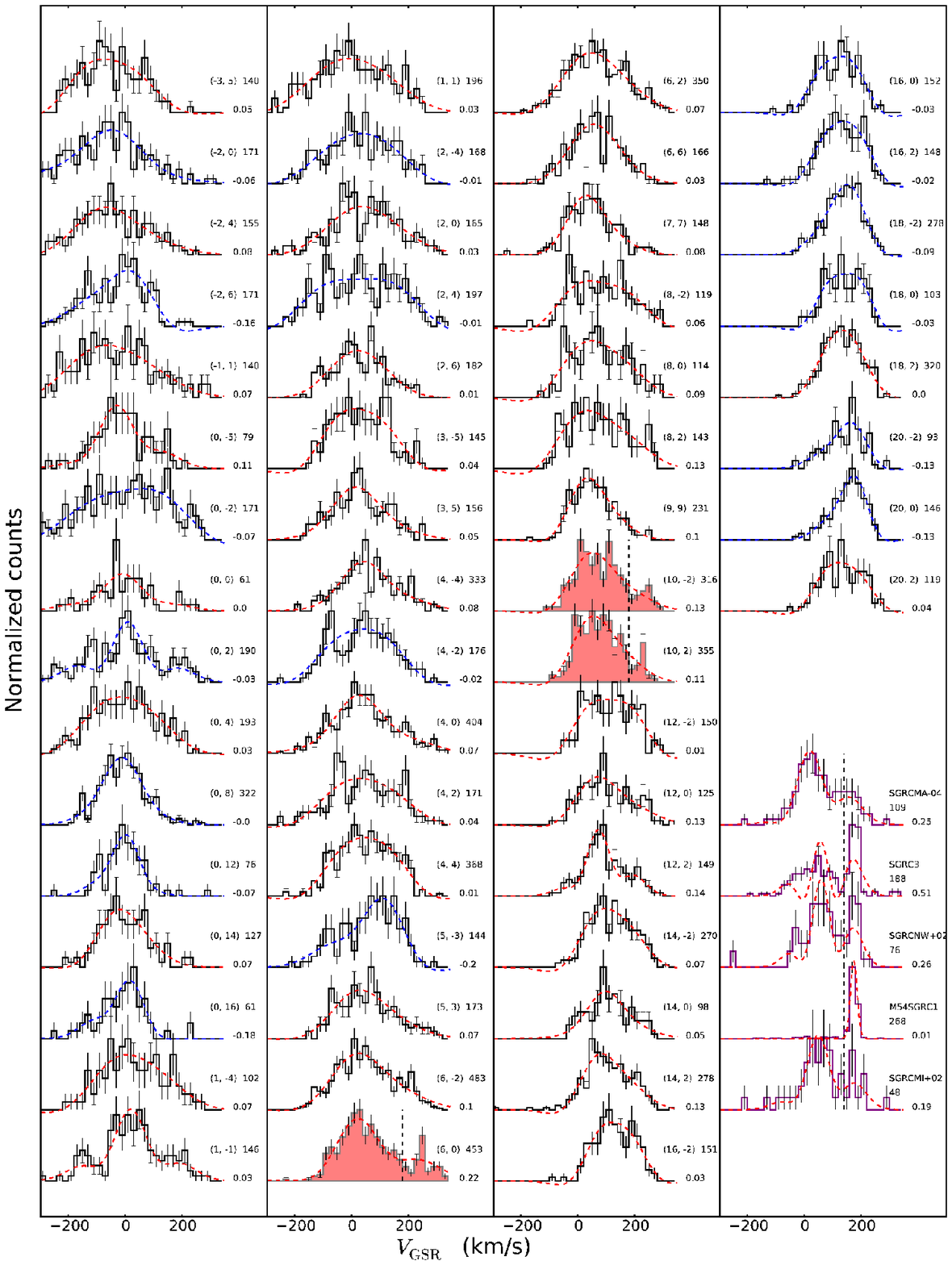}
\caption{ Velocity distributions in the fields within $-5^{\circ} < l < 20^{\circ}$ and $-20^{\circ} < b < 20^{\circ}$. The red filled histograms represent the three fields showing strong HV peaks at a similar velocity. The purple histograms to the lower right are the Sgr fields.  For each distribution, the best-fit Gauss$-$Hermite polynomial is overplotted, with red and blue curves representing fits resulting in positive and negative $h_{3}$ values, respectively. The field position ($l$, $b$) (upper left), the number of stars (upper right), and the $h_{3}$ value (lower) are shown beside each profile. In the three red histograms and the Sgr fields, the vertical dashed lines represent the separation between the HV peaks and the main component.}
\label{fig-vg-all}
\vspace{0.2cm}
\end{figure*}

\begin{figure*}[!t]
\centering
\includegraphics[width=1.6\columnwidth]{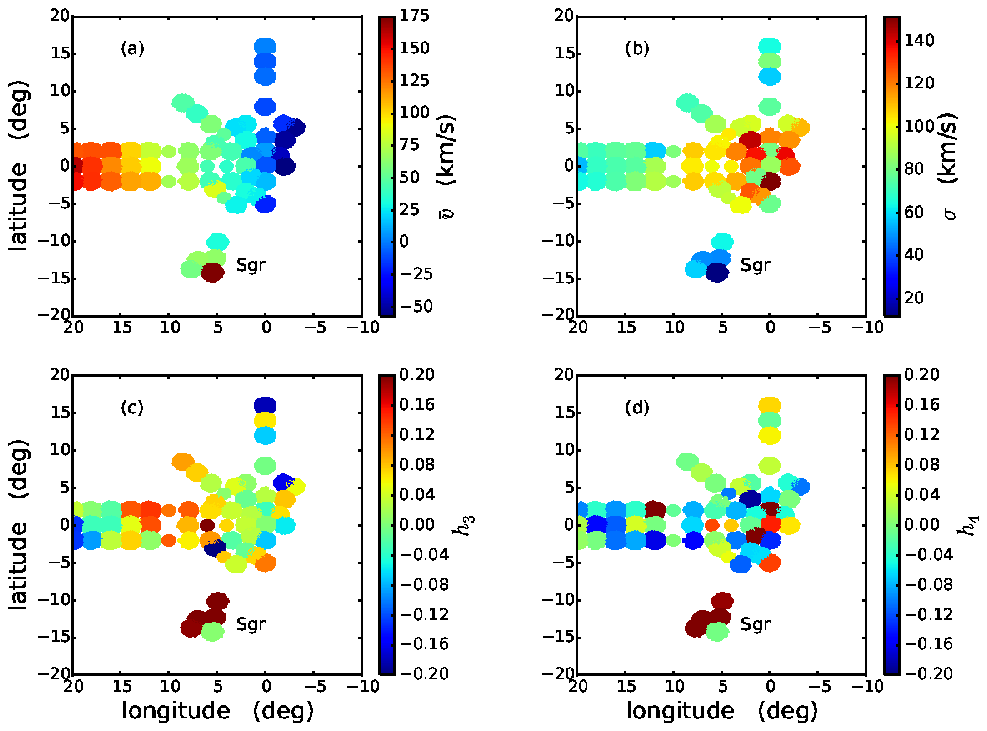}
\caption{Angular distribution of the measured velocity distribution parameters for APOGEE DR13 stars in the Galactic bulge region, color-coded according to the values of the four parameters of Equation \ref{Gauss_Hermite}, i.e. , (a) $\bar{v}$, (b) $\sigma$, (c) $h_{3}$, and (d) $h_{4}$.}
\label{h3h4}
\vspace{0.2cm}
\end{figure*}

Figure \ref{fig-vg-all} displays the velocity distributions for all $-5^\circ < l < 20^\circ$ and $-20^\circ < b < 20^\circ$ survey regions. The bin width of 20 $\rm km\,s^{-1}$ is the same as in \cite{Nidever2012}. Clear HV peaks can be seen in only three bulge fields, i.e., at ($l$, $b$)=(6$^\circ$, 0$^\circ$) and (10, $\pm 2$)$^\circ$. For this discussion we only consider positive HV peaks centered at 150-220 $\rm km\,s^{-1}$, where the distributions are clearly skewed Gaussians from visual inspection. Peaks at negative velocities such as the field (4$^\circ$, $-$2$^\circ$) are not considered because no current theoretical models can explain such peaks. A simple statistical approach\footnote{We define a statistic $s=|n-y|/\sqrt{n}$, where $n$ is the number of counts in a certain bin and y is the theoretical value from single Gaussian fitting. The value $\sqrt{n}$ is the Poisson error. Our null hypothesis is that the peak is significant. The confidence level could be calculated by $\rm erf(\frac{s}{\sqrt{2}})$, where $\rm erf(x)$ is the error function.} is adopted to estimate the significance of the peaks compared to the Poisson noise. Most negative velocity peaks and some of the positive peaks might be due to the statistical fluctuations because they are within  $\sim2\sigma$ Poisson error. However, at (6$^\circ$, 0$^\circ$), (10$^\circ$, $+2$$^\circ$) and (10$^\circ$, $-2$$^\circ$) the statistic values of the peaks are 4.90$\sigma$, 3.77$\sigma$, and 2.88$\sigma$, respectively. As a comparison, in Section \ref{result_with_disk}, we also present velocity distributions of our sample without removing foreground disk stars, and the majority of fields show only a smooth skewed Gaussian distribution, and the peaks become statistically less significant.

To compare directly with \cite{Nidever2012} in the radial velocity distributions for different bulge fields, we use the same bin size (20 $\rm km\,s^{-1}$) as in their paper.  Different bin sizes can affect the shape of the distribution: larger bins reduce the sampling noise, while smaller bins give better resolution of the density estimation but with larger noise. The optimal choice of the bin size for the histogram  determined with the Freedman \& Diaconis' rule\footnote{\cite{Freedman1981} specifies an optimal method to minimize the difference between the histogram and the underlying distribution. The optimal bin size is  $h=2IQR(X)n^{-1/3}$, where $n$ is the number of observations on $X$. $IQR(X)$ is the interquartile range (the difference between the upper [top 75 \%] and lower [bottom 25 \%] quartiles.}
is from  25 $\rm km\,s^{-1}$ to 70 $\rm km\,s^{-1}$ (depending on the number of stars in each field), which is larger than the adopted bin size 20 $\rm km\,s^{-1}$.
With the optimal bin sizes, potential HV peaks all become weaker or even disappear. Therefore, statistical fluctuation due to small sample sizes might also be partially responsible for the HV peaks, as suggested by \cite{Lizhaoyu2014}.

The underlying distribution of velocity is still unknown. A double Gaussian is the simplest model, as attempted in \cite{Nidever2012} and \cite{Zasowski2016}. In this case, the positions of the peaks are very important parameters. However, if the peak position is allowed to vary during the fitting, it is quite difficult to constrain the two Gaussians simultaneously (one for the main component, the other for the HV peak), i.e., two different double Gaussian profiles might give very similar $\chi^2$. On the other hand, if the peak positions of the double Gaussian are constrained in a small range during the fitting, the final results would depend sensitively on the initial peak positions. Therefore this fitting strategy may not be optimal. There is another way to describe a skewed Gaussian. \cite{vanderMarel1993} proposed a decomposition of line profiles into orthogonal functions of the Gauss$-$Hermite series. The LOSVD could be written as \citep{vanderMarel1993}:

\begin{equation}	
	LOSVD(v)=\frac{\gamma}{\sqrt{2\pi}\sigma}e^{-(v-\bar{v})^2/2\sigma^2}(1+h_{3}H_{3}(\frac{v-\bar{v}}{\sigma})+h_{4}H_{4}(\frac{v-\bar{v}}{\sigma})),
\label{Gauss_Hermite}
\end{equation}	
where $\bar{v}$ is the mean velocity, $\sigma$ is the velocity dispersion, $H_{3}(\omega)=\frac{1}{\sqrt{3}}(2\omega^3-3\omega)$, and $H_{4}(\omega)=\frac{1}{\sqrt{24}}(4\omega^4-12\omega^2+3)$. In this function, $\gamma$, $\bar{v}$, $\omega$, $h_{3}$, and $h_{4}$ are free parameters. We truncate the Gauss$-$Hermite series at $n=$ 4 because the higher-order moment contribution is negligible \citep{Galactic_Astronomy1998}.
The parameters $h_{3}$ and $h_{4}$ measure asymmetric and symmetric deviations from a Gaussian, respectively. For example, a distribution with a prominent HV tail will have a positive $h_{3}$ and a distribution with a sharp central peak will have a positive $h_{4}$.

To avoid the uncertainties and degeneracies mentioned above, and to better describe the shape of the velocity distribution, the Gauss$-$Hermite series are used in this paper (as in \citealt{Bureau2005, Iannuzzi2015, Dumin2016}). Considering the complicated bar kinematics and our small sample, Gauss$-$Hermite moments that describe higher order deviations of a Gaussian may be a better choice than fitting double Gaussian profiles.

The Gauss$-$Hermite fitting results are shown in Figure \ref{fig-vg-all} and the corresponding $\bar{v}$, $\sigma$, $h_{3}$ and $h_{4}$ distributions are shown in Figure \ref{h3h4}. 
Velocity distributions in $5^{\circ} < l < 15^{\circ}$ and $-3^{\circ} < b < 3^{\circ}$ show larger $h_{3}$ than other fields. Including the three fields with clear HV peaks, the velocity distributions in this region  show strong asymmetries. These will be studied in more detail in Section 4 specifically for their chemical properties.

Although the velocity distributions look noisy, the HV shoulders in most of the bulge fields can be well described by the Gauss$-$Hermite polynomials, except the three fields showing clear HV peaks. We will show that the spatial distribution of $h_{3}$ is consistent with that of the model in \cite{Shen2010} (Section \ref{section3.2}) and a clearer pattern of $h_{3}$ is seen when we do not remove the foreground disk giants (Section \ref{result_with_disk}).

\cite{Debattista2015} suggested that the HV peaks of the LOSVD could be due to a kiloparsec-scale stellar disk that is composed of stars on $x_{2}$ orbits. They predict that the LOSVDs in the midplane ($l=8^{\circ}\pm2^{\circ}$) should exhibit HV peaks, whereas those off-plane ($|b| \ge 1^{\circ}$) will not. The clear peak at (6$^\circ$, 0$^\circ$) may be explained by their model provided that the nuclear disk/ring is smaller. However, the peaks off-plane at (10$^\circ$, $\pm$2$^\circ$) cannot be explained by their model and remain puzzling in the context of their model.

APOGEE has fields covering the Sgr core and stream. In these fields candidate Sgr members are targeted based on a selection of 2MASS M giants \citep{Majewski2013, Zasowski2013}.
Since the Sgr has experienced quite different evolution history compared to the MW, the comparison between the SGR field and Galactic bulge is useful to distinguish the chemical and kinematic differences. In Sgr fields, especially SGRC3, the HV peaks are much more prominent; thus, the fitting result is not as good as that in the bulge fields. This implies that the underlying velocity distribution is different from that of bulge stars. The HV peaks in Sgr fields are better described by adding another Gaussian. This is reasonable since the stars in the second peak are the core or tidal streams of the Sgr dSph. In this case, a double Gaussian fit may be more appropriate.

\subsection{$h_{3}$ and skewness profiles in the Galactic bulge}
\label{section3.2}
\begin{figure*}[!t]
\centering
\includegraphics[width=1.8\columnwidth]{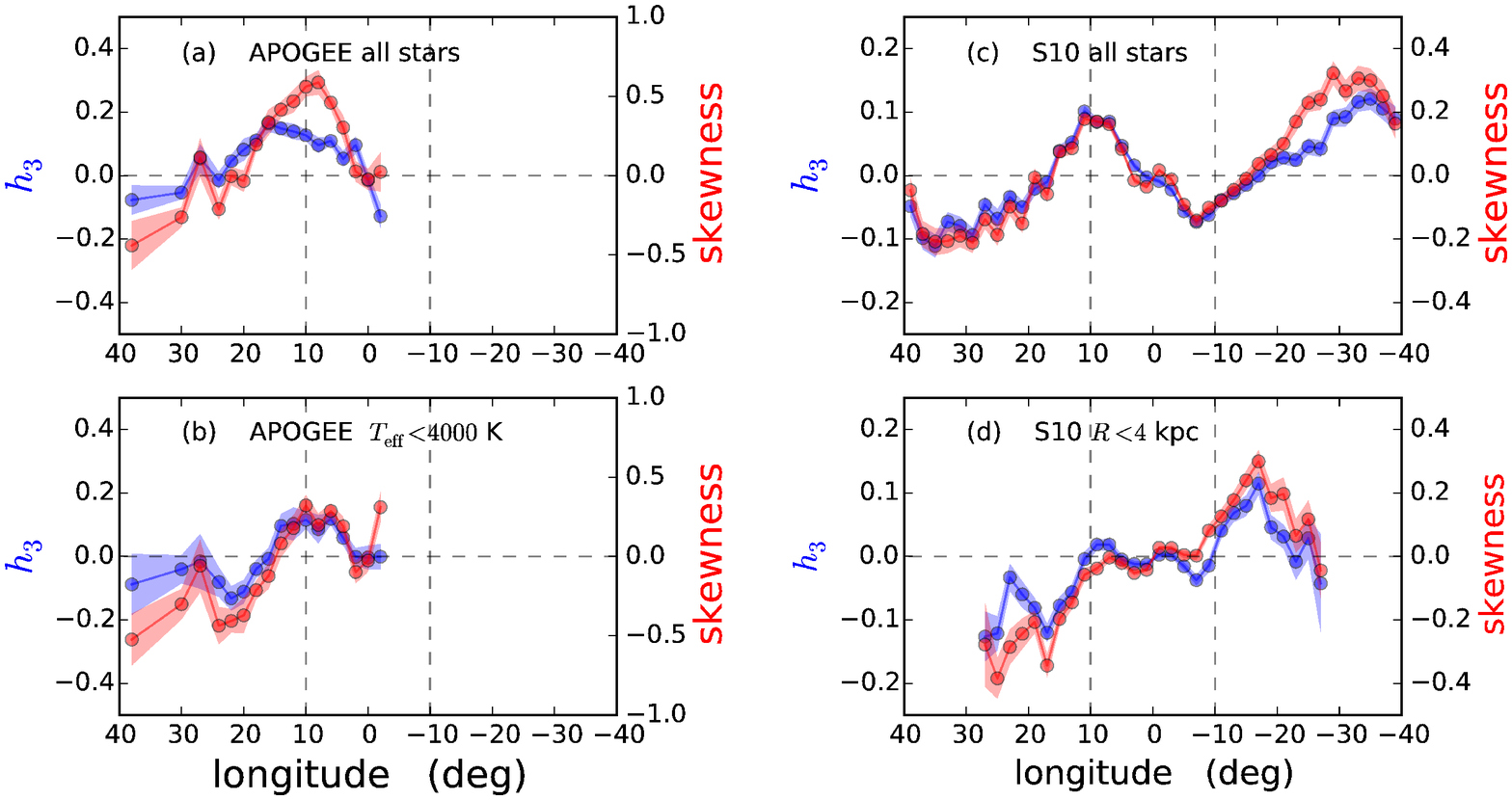}
\centering\caption{The $h_{3}$ (blue) and skewness (red) profiles as a function of the Galactic longitude (a) for APOGEE stars without foreground stars excluded; (b) for APOGEE stars with foreground stars excluded, using $T_{\rm eff}$ as a distance proxy; (c) for all particles of model S10; and (d) for particles with distances less than 4 kpc from the Galactic center of model S10. The 1$\sigma$ errors derived from the bootstrapping method are shown as blue and red shaded regions in each panel. The vertical dashed lines mark the boundary of the bulge region. Note that the mean velocity profile ($\bar{v}$) increases monotonically from negative to positive $l$, as seen in Figure \ref{h3h4}(a).}
\label{4plot}
\end{figure*}

Simulations suggest that $h_{3}$ provides a good description of the shape of LOSVDs in the bar region of edge-on galaxies.  In an edge-on disk, for a bar seen end-on\footnote{ End-on means that the bar is viewed in-plane from its major axis.}, $h_{3}$ is correlated with $\bar{v}$ over the entire projected bar length, and then becomes anticorrelated beyond the bar in the disk region, and becomes correlated again at even larger distances from the galaxy center \citep{Bureau2005}. This feature makes an $h_{3}$-$\bar{v}$ correlation an indication of the LOSVD with the HV tail created by the bar-supporting orbit \citep{Shen2009, Iannuzzi2015}. Thus, $h_{3}$ can be a very good tracer of bars viewed edge-on, instead of using a second peak.
The correlation between $h_{4}$ and the bar is not as strong as $h_{3}$ but we consider it for completeness. For the strong-bar case, the width of the central $h_{4}$ minimum is roughly the same as that of the flat part of the dispersion profile \citep{Bureau2005}.

Skewness and kurtosis\footnote{The skewness is the third standardized moment, defined as $g_{1} = m_{3}/m_{2}^{3/2}$, where $m_{3} = \Sigma(x-\bar{x})^3/n$ and $m_{2} = \Sigma(x-\bar{x})^2/n$, $\bar{x}$ is the mean, and $n$ is the sample size. The sample skewness we use is $(n(n-1))^{1/2}/(n-2)\times g_{1}$ \citep{Joanes1998}. Kurtosis is the fourth standardized moment, defined as $g_{2} = m_{4}/m_{2}$.} are also used to described the bar kinematics \citep{Zasowski2016}. Fitting the full LOSVD requires all the moments of the distribution. Higher-order moments depend sensitively on the LOSVD shape, especially the contribution from outliers. The coefficients $h_{3}$ and $h_{4}$ of the Gauss$-$Hermite series can give a good description of the skewness and kurtosis, with additional restrictions on the shape of the underlying distribution.  Considering that the covariances between the best-fit parameters are minimized, using Gauss$-$Hermite polynomials is optimal. In other words, there will be almost no correlations between the errors in different parameters \citep{vanderMarel1993}.

To explore how $h_{3}$ changes with $l$ (or $\bar{v}$), we compare the observational data to an $N$-body model of the MW bar. The model employed here comes from \citet[hereafter S10]{Shen2010} which has successfully reproduced MW bulge kinematics from BRAVA observations. The simulation starts with an exponential disk represented by one million particles in a rigid dark matter halo potential. A bar forms spontaneously owing to disk instability and quickly develops a boxy/peanut-shaped bulge in the vertical direction. We select the snapshot at $\sim$5 Gyr to study the bulge kinematics here. The half length of the bar in S10 is 4 kpc, and the Sun-Galactic center line is $20^\circ$ offset from the major axis of the bar.

Figure \ref{4plot} shows the $h_{3}$ and the skewness profiles as a function of the longitude in our sample ($-5^\circ<l<40^\circ$, $-3^\circ<b<3^\circ$) and the simulation from S10 ($-40^\circ<l<40^\circ$, $-3^\circ<b<3^\circ$). We use the bootstrap method to estimate the errors of the sample. The 1$\sigma$ errors are shown as blue and red shaded regions in each panel. We found that the skewness measurement can show large scatter owing to the velocity outliers dominating the $(v-\bar{v})^3$ term in the skewness computation, while $h_{3}$ is much less affected in this case. Therefore, we exclude stars with $\rm V_{GSR}$ deviating from $\bar{v}$ by 3$\sigma$ to avoid large fluctuations in the skewness distributions.

In S10, with all the particles, the bulge (inner) and the disk (outer) regions show opposite behaviors in both $h_{3}$ and skewness profiles, as shown in Figure \ref{4plot}(c). Note that the mean velocity profile ($\bar{v}$) increases monotonically from negative to positive $l$ (see Figures \ref{h3h4}(a) and Figure \ref{model}). Inside the bulge region ($|l| < 10^\circ$ ),  $h_{3}$ and skewness have positive correlation with the line-of-sight velocity for the model, while in the disk-dominated region ($|l| > 10^\circ$ ), the correlation changes to negative (anticorrelation). For a pure disk dominated by circular motions, the velocity distribution should have a low-velocity tail (negative $h_{3}$) since along each line of sight the tangent point contributes the largest fraction of stars and the largest velocity, shifting the peak of the LOSVD toward the HV direction with an extended low-velocity tail. For a bar viewed nearly end-on, along each line of sight, the bar component would contribute a large fraction of HV stars, resulting in a positive $h_{3}$ and skewness in LOSVD \citep{Bureau2005}.

In Figure \ref{4plot}(d), we exclude the particles with distance larger than 4 kpc away from the Galactic center to select only those in the bulge without foreground or background contaminations. The $h_{3}-\bar{v}$ correlation in the bar region in Figure \ref{4plot}(d) becomes much more weakened compared to that in Figure \ref{4plot}(c).  At positive longitudes, the foreground and background stars mainly contribute to low-velocity tails. With them removed, the velocity distribution becomes roughly symmetric to yield small $h_{3}$ and skewness values. Away from the bulge region, they become negative since disk stars dominate the LOSVD. The behaviors are reversed for negative longitudes.
For our APOGEE sample, with all stars considered, the $h_{3}$ and the skewness show similar trends, but the $h_{3}$  peaks at $l \sim$ 17$^\circ$ while the skewness peaks at  $l \sim$ 10$^\circ$. After excluding the foreground disk stars  as shown in Figure \ref{4plot}(b), both the $h_{3}$ and the skewness  peak at $l \sim$ 10$^\circ$. 
The observational results with all the stars considered are roughly consistent with the S10 model prediction, although the exact values are different. Figure \ref{4plot}(b) shows larger $h_{3}$ and larger skewness than Figure \ref{4plot}(d) in the bulge region in the range $l \sim 3^\circ - 12^\circ$. There are two possibilities to explain this difference. First, there might still be foreground disk stars left in our sample; the foreground stars mainly contribute to the low-velocity distribution in the LOSVD, resulting in an HV tail and positive $h_{3}$ value. Second, this may reflect that there are clear HV peaks in some fields that cannot be explained by the S10 model. The simulation may lack the HV stars to generate such a feature. Our results here are broadly consistent with the independent skewness analysis in \cite{Zasowski2016}. 

\subsection{Results including foreground stars}
\label{result_with_disk}

\begin{figure*}[!t]
\centering
\includegraphics[width=1.8\columnwidth]{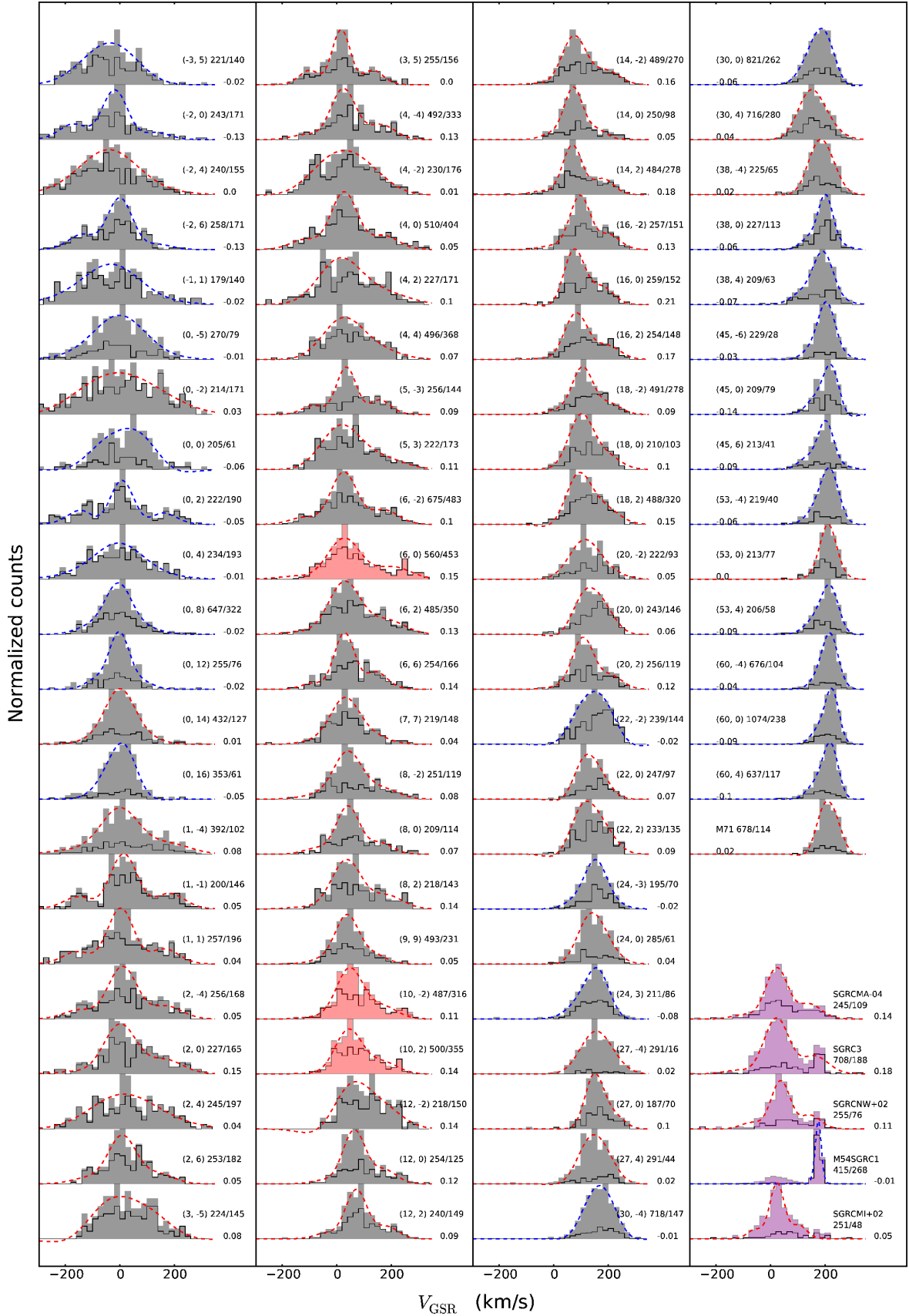}
\centering\caption{Velocity distributions of the APOGEE fields in the range $-5^{\circ} < l < 60^{\circ}$ and $-20^{\circ} < b < 20^{\circ}$ without foreground disk star exclusion. The purple histograms are for the Sgr fields.  For each distribution, the best-fit Gauss$-$Hermite polynomial is overplotted, with red and blue curves representing the positive and negative $h_{3}$ values, respectively. Note that the clear peak at (10$^\circ$, $-$2$^\circ$) in Figure \ref{fig-vg-all} becomes less striking in this figure. The black lines show the velocity distribution with foreground stars excluded. The field position ($l$, $b$) (upper left), the number of stars before/after foreground cut (upper right), and the $h_{3}$ value (lower) are shown beside each profile.}
\label{vg_fit_all}
\end{figure*}

In our study we first remove the foreground stars in the bulge region using a $T_{\rm eff}$ cut. The results presented here should be less contaminated by disk stars than those in previous APOGEE studies (e.g., \citealt{Nidever2012}; \citealt{Zasowski2016}). However, as suggested by \cite{Zasowski2016}, with the foreground stars included, the results better mimic integrated light "observations" of the MW. Therefore, we also present the bulge kinematics without foreground stars removed. Figure \ref{vg_fit_all} shows the LOSVDs of the fields in the region defined by $-5^{\circ} < l < 60^{\circ}$ and $-20^{\circ} < b < 20^{\circ}$ . The color scheme is the same as in Figure \ref{fig-vg-all}. 
Most of the fields only show smooth HV shoulders. Note that there are almost no stars removed from the HV bins, which means that HV stars live in the bar region, although they need not be bar stars.

\begin{figure*}[!t]
\centering
\includegraphics[width=1.8\columnwidth]{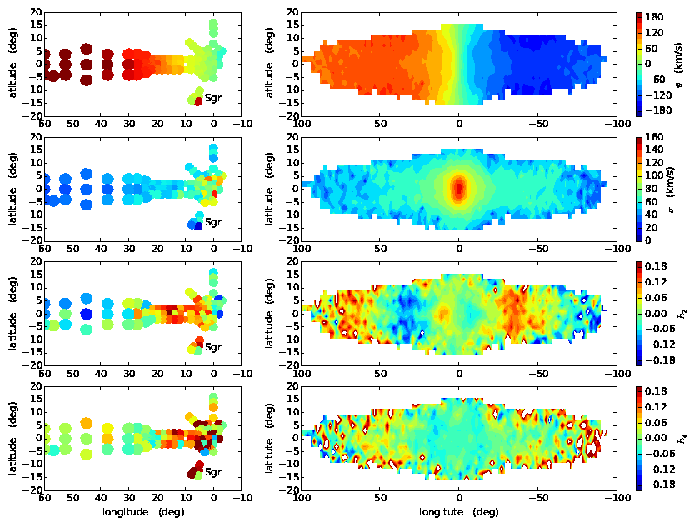}
\centering\caption{Left panels: $\bar{v}$, $\sigma$, $h_{3}$, and $h_{4}$ maps of all APOGEE stars without removing foreground disk stars in the coordinate range $-5^{\circ} < l < 60^{\circ}$ and $-20^{\circ} < b < 20^{\circ}$. Right panels: $\bar{v}$, $\sigma$, $h_{3}$, and $h_{4}$ maps of the S10 simulation of the Galactic bar. }
\label{model}
\end{figure*}

The average velocity, $\sigma$, $h_{3}$ and $h_{4}$ maps are shown in the left column of Figure \ref{model}. In the bulge region, $\bar{v}$ shows clear cylindrical rotation. As the longitude increases, $\bar{v}$ increases monotonically, while $h_{3}$ increases and reaches maximum value at $l \sim 15^{\circ}$ and then decreases in the disk-dominated region ($l>20^{\circ}$). Apparently, $\bar{v}$ and $h_{3}$ show positive correlation inside the bulge region and negative correlation in the disk part. This is consistent with the simulation predictions (\citealt{Bureau2005}, \citealt{Shen2009}, \citealt{Iannuzzi2015}). The $h_{3}$ and $h_{4}$ maps are consistent with the skewness and kurtosis maps derived in \cite{Zasowski2016}.

For comparison, we also derive the kinematic maps for the S10 model with all the particles considered, as shown in the right column of  Figure \ref{model}. The general trend in $\bar{v}$ and $h_{3}$ is quite similar with our observational results. However, for the model in the bulge fields, $h_{4}$ is generally quite small. This test gives further confirmation that the MW bulge/bar model provides a good match to the data, except for $\sigma$ and $h_{4}$, which look different from the observation. In the bulge fields, the positive $h_{4}$ values in the observation indicate a significant central narrow peak, which may be contributed mainly by the foreground disk stars. The disk in the S10 model may not be cold enough to generate such a central narrow peak. 
We also calculated the velocity dispersion directly as the sample standard deviation and found good agreement between the data and model (see also Figure 3 in \citealt{Zasowski2016}). The lower fitted $\sigma$ in the data may be due to the shape of LOSVDs which have higher $h_{4}$ values. This is also suggested by the relation between the standard deviation $\tilde{\sigma}$ and the Gauss$-$Hermite best-fit parameter $\sigma$ ($\tilde{\sigma} \approx \sigma(1+\sqrt{6}h_{4})$; Equation 18 in \citealt{vanderMarel1993}).

\section{Chemical properties of the bulge stars}
\subsection{Chemical abundances of the HV fields}

\begin{figure*}[!t]
\centering
\includegraphics[width=1.8\columnwidth]{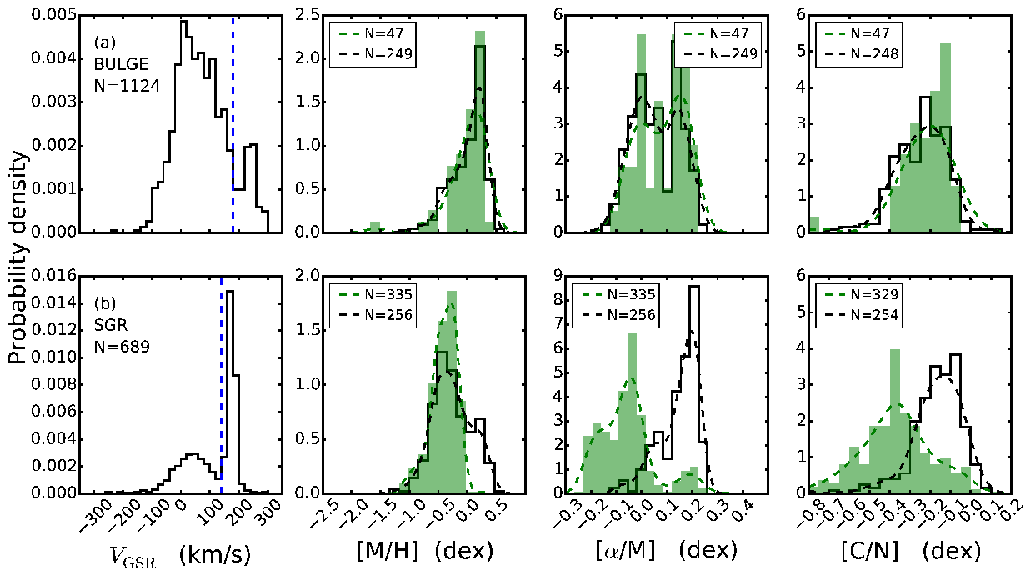}
\caption{Probability density function of velocity, [M/H], [$\alpha$/M] and [C/N] of the combined three bulge fields (top row) showing clear HV peaks, i.e., (6$^\circ$, 0$^\circ$) and (10$^\circ$, $\pm$2$^\circ$), and the Sgr fields (bottom row). 
In the left panels, the vertical dashed lines mark the separation of the HV peaks and the main component. In the right three columns, the black histogram shows the main component distribution, while the green filled histogram shows the distribution of the HV peaks. For each panel, the corresponding KDE profiles using Silverman's rule are also shown with green and black dashed lines.}
\label{fig:2fields}
\vspace{0.2cm}
\end{figure*}

In this section we study the stellar chemical properties. We use the chemical abundances determined by ASPCAP \citep{Garcia2016} and use ASPCAPFLAG to select our sample. Stars with any bad flags or warnings on [M/H], [$\alpha$/M], [C/Fe] and [N/Fe] are excluded. It would be reasonable to expect that different stellar populations might have different chemical abundances. The (6$^\circ$, 0$^\circ$) and (10$^\circ$, $\pm$2$^\circ$) fields show large $h_{3}$ values, indicating strong asymmetric deviations in the velocity distribution. The HV peaks in the three fields are significant at a confidence level of 0.996, which is unlikely as a result of Poisson noise. If the HV peak really exists, focusing on these three fields would be more physically meaningful. Comparing the chemical distributions between HV peaks and the main component in the three fields may reveal the mechanism behind the HV peaks. We also include the five Sgr fields as a control sample for this analysis.

 The velocity distributions of these fields are shown in the first column of Figure \ref{fig:2fields} with (6$^\circ$, 0$^\circ$) and (10$^\circ$, $\pm$2$^\circ$) combined together in the top panels and the SGR fields in the bottom panels. Stars are divided into two components with a velocity cut at 180 $\rm km\,s^{-1}$ (in the three bulge fields) and 100 $\rm km\,s^{-1}$ (in the five Sgr fields).
The chemical abundance distributions including metallicity, [$\alpha$/M] and [C/N] of these fields are shown in the other three columns of Figure \ref{fig:2fields}. For each histogram the kernel density estimation (KDE) is also shown, where the Silverman's rule\footnote{Silverman's rule \citep{Silverman1986} gives the optimal kernel that would minimize the mean integrated squared error if the data were Gaussian or a Gaussian kernel is  used. The optimal bandwidth (kernel) $h$ is:  $h=1.06 A {n^{-1/5}} ,  A=min(s,IQR(X)/1.34)$,  where $n$ is the number of observations on $X$, $s$ is the standard deviation of the sample, and $IQR(X)$ is the interquartile range (the difference between the upper [top 75 \%] and lower [bottom 25 \%] quartiles).} is used to determine the kernel size.

The metallicity distributions of the main component and the HV peaks in the three bulge fields are very similar in terms of the range, peak position, and shape of the distribution. In the (6$^\circ$, 0$^\circ$) field, \cite{Babusiaux2014} found that the HV stars are $\sim$ 0.1 dex higher in [Fe/H] than the main component. However, the difference is smaller in our result, where the two components have almost the same mean metallicity \textbf (HV stars are $\sim$ 0.002 dex higher). 
 In contrast, in the Sgr fields the metallicity distribution of the HV peak is quite different from the rest of the stars. The HV stars in the Sgr field are more metal-poor with smaller scatter.

In addition, the [$\alpha$/M] distributions of the HV peaks are quite similar to that of the main components in the three bulge fields. Moreover, both the main components and HV peaks display bimodal distributions, which is a strong indication of multiple stellar populations. In the Sgr field, the HV stars are clearly less $\alpha$-enhanced, suggesting different stellar populations and star formation history between the stars in the HV peak (the core) and the main component (the foreground stars of MW) in Sgr fields.

As shown in Figure \ref{fig:2fields}, the [C/N] distributions of the HV peaks and the main component in the three bulge fields are also similar. On the other hand, in the Sgr field the HV peak and the main component show clear differences: the scatter of HV stars is larger and the mean value of the HV peak is smaller than that of the main component. 

For a more quantitative comparison, we also apply the K$-$S test on the distributions, where the statistic $p$ describes the significance of the difference between the two samples. If $p$ > 0.05, we accept the null hypothesis, which means that there is no difference between the two samples. We test the chemical abundance distribution for all fields in the region $5^{\circ} < l < 15^{\circ}$ and $-3^{\circ} < b < 3^{\circ}$. As expected, the K$-$S tests show that [M/H], [$\alpha$/M] and [C/N] of the two components follow the same distribution in most of these fields.  

Although the uncertainties of stellar parameters for cooler stars ($T_{\rm eff}$ < 4000 K) are statistically larger \citep{Holtzman2015}, the difference between the two components is still very significant in the chemical abundance for Sgr fields. For example, in the Sgr field, there are clear differences between the HV and the main components in the metallicity distributions. This is expected because the HV peaks are contributed mainly by the member stars of the Sgr dSph. This demonstrates that the uncertainties are probably not large enough to blur the distribution of different stellar populations. Therefore, the similarities in the chemical distributions between the main component and the HV peaks in bulge fields suggest similar stellar populations.

\subsection{Age distinctions in bulge fields}

\label{1234}
\begin{figure*}[!t]

\begin{minipage}{0.48\linewidth}
\centerline{\includegraphics[width=1\textwidth]{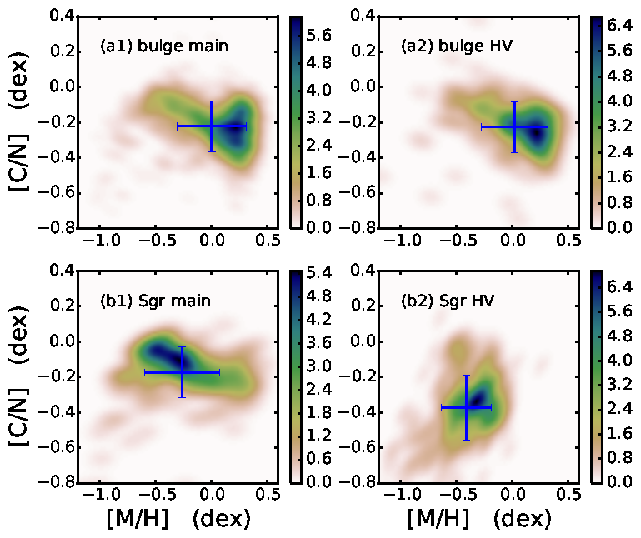}}

\centerline{}
\end{minipage}
\qquad
\begin{minipage}{0.48\linewidth}
\centerline{\includegraphics[width=1\textwidth]{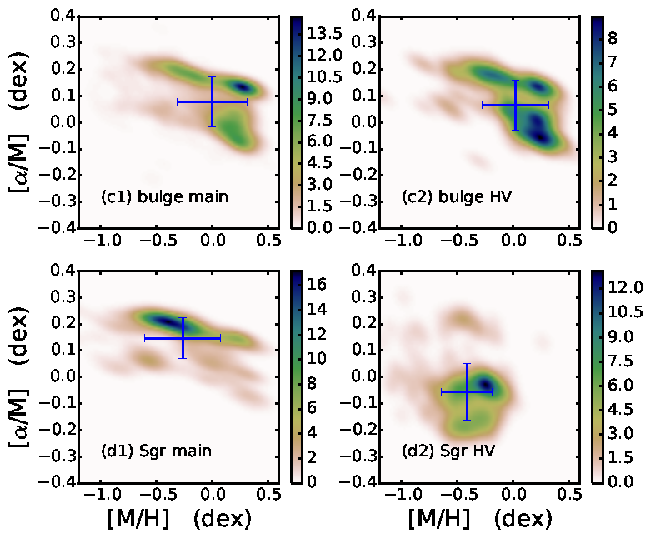}}
\centerline{}
\end{minipage}

\caption{KDE smoothed distributions of stars in the bulge fields between $5^{\circ} < l < 15^{\circ}$ and $-3^{\circ} < b < 3^{\circ}$ (panels (a) and (c)) and the Sgr fields (panels (b) and (d)) in [M/H]$-$[C/N] (panels (a) and (b)) and [M/H]$-$[$\alpha$/M] (panels (c) and (d)). The mean value and standard deviation are also shown in this figure. The stars are further divided into main component (1) and HV component (2) based on their velocity. The difference  between the two components can be clearly seen in Sgr fields, but not in the bulge.}
\label{fig_cnammh}

\end{figure*}

Directly measuring the stellar age from observations is nontrivial. Metallicity and $\alpha$-enhancement can only provide rough estimates of age. Recently, \cite{Martig2015} argued that the relative surface abundances of C and N of RGB stars that have experienced the first dredge-up process can also be a good age indicator. The first dredge-up process brings material produced by the CNO cycle from the bottom of the convection layer to the surface. The amount of N brought to the surface depends on their initial masses. For more massive stars, the fraction of nitrogen in their
cores is higher and the convective zone extends deeper. After the dredge-up, the surface abundance of [(C+N)/M] is unchanged since the total amount of CNO is conserved and the abundance of O is only slightly affected by the dredge-up. As a result, more massive stars have higher surface abundance of N and lower surface abundance of C, leading to a lower [C/N] ratio compared to low-mass stars \citep{Martig2015}.
Lower-mass stars live longer during the main-sequence phase, and hence are older on average when they proceed to the RGB phase, and vice versa for the high-mass stars.  Therefore, [C/N] for RGB stars can be a good indicator of stellar age  \citep{Martig2015}; lower (higher) [C/N] values correspond to younger (older) stars. Figure 5 in \cite{Martig2015} shows the empirical relation of [M/H], [C/N] and median age, with a median age scatter of 26 $\%$. Note that the chemical abundances we use are from the same pipeline. The uncertainty of [C/N] is about 0.1 dex \citep{Masseron2015, Martig2015}.

The three right columns of Figure \ref{fig:2fields} show different proxies of age. The statistically identical distributions between the main component and the HV peaks for the bulge fields (see Section 4) suggest that the two components seem to have the same stellar populations. Note that [C/N] as an age proxy is only valid for RGB stars, not for asymptotic giant branch (AGB) stars. The contamination from AGB stars may pollute the result. However, given that AGB stars have a much shorter lifetime compared to stars on the RGB, at least statistically there are far fewer AGB stars in the sample than RGB stars.

Inspired by \citet{Martig2015}, we use the multivariate distribution of [M/H] and [C/N] to further investigate  the potential age differences, since stars with different ages are located at different positions in the [M/H]$-$[C/N] plane. Based on their velocity, stars in the range $5^{\circ} < l < 15^{\circ}$ and $-3^{\circ} < b < 3^{\circ}$ are divided into two components: the main component ($V_{GSR}$ < 180 $\rm km\,s^{-1}$) and the HV component ($V_{GSR}$ > 180 $\rm km\,s^{-1}$).  The SGR stars are also defined, but with a different velocity cut (140 $\rm km\,s^{-1}$).
Figure \ref{fig_cnammh} shows the density distributions on [M/H]$-$[C/N] and [M/H]$-$[$\alpha$/M] planes for different velocity components in the bulge and in the Sgr. The mean value and the standard deviation are also shown in the figure. 

The metallicity of bulge stars (both the main component and the HV component) is mainly from $-$1.2 to 0.5 dex, while [C/N] is from $-$0.6 to 0.2 dex. From Figure \ref{fig_cnammh} the [C/N] of metal-rich stars spreads over a wide range. Since [C/N] is a good age proxy that does not depend on star formation histories, a wide spread of [C/N] indicates a wide age range. \citet{Bensby2013} studied the chemical abundance of 58 microlensed bulge dwarfs and subgiants and found that metal-rich stars have a wide age distribution with a large fraction of young and intermediate-age stars. Using [C/N] as an age proxy, our result is consistent with that by \citet{Bensby2013}. Compared with \citet{Martig2015}, the age of the metal-rich stars (0 < [M/H] < 0.4 dex) mainly ranges from 2 to 14 Gyr. The [C/N] values of the metal-poor stars also spread over a wide range, but most of them are concentrated from $-$0.2 to 0 dex. The higher [C/N] indicates that these stars are older (6$-$14 Gyr according to the [M/H]$-$[C/N] distribution of \citealt{Martig2015}) than the metal-rich stars\footnote{\cite{Martig2015} established the empirical relationship of [C/N], [M/H] and age for upper RGB stars with $T_{\rm eff} >$4000 K. Our sample contains stars cooler than 4000 K. We cannot fully exclude that stars with the same [C/N] might have different age ranges at different temperatures. This relation could be used if stars have experienced their first dredge-up phase. Most of the stars in our sample are red giants with $\rm log$$\;g<3.8$. We also find that stars of different temperature in the bulge region follow the same distribution of [M/H], [C/Fe] or [N/Fe]. So we infer that the temperature may not affect the empirical relation between [C/N] and age.}. 

On the [M/H]$-$[$\alpha$/M] plane, the [$\alpha$/M] values of the two samples seem to spread over the same range and have similar mean value and standard deviation. 
The relative fraction of low-$\alpha$ stars in the HV component is slightly higher than that in the main component. We suspect that this is due to the ambiguous separation of the HV peak. In Figure \ref{fig:2fields}, which separates the HV stars much better than Figure \ref{fig_cnammh}, the $\alpha$-enhancement of the HV peak stars shows almost an identical distribution with that of the main peak stars. Of course, $\alpha$-enhancement is not as good an age indication as [C/N].

The $\alpha$-enhanced branch in [M/H]$-$[$\alpha$/M] distribution indicates a rapid formation scenario, while the low-$\alpha$ branch indicates a longer formation time scale. Note that \cite{Johnson2014} did not find a bimodal [M/H]$-$[$\alpha$/M] distribution for stars in the bulge region. We suspect that the contamination of the foreground disk might contribute to the observed bimodal distribution here.

By using simulations in which newly formed stars are considered, \cite{Aumer2015} suggested that the selection function of APOGEE is biased to young stars and that the HV feature is preferentially composed of young stars less than 2 Gyr old. 
Thus, we could speculate that the average age of the HV component is younger than the main component. However, if the age compositions of the two components are different, this should be visible from the chemical abundance distributions. We can infer from Figures \ref{fig:2fields} (a), \ref{fig_cnammh} (a1), and (a2) that the age compositions of the two velocity components are similar and the fractions of young stars are low.

As shown in \cite{Bensby2017}, the age-metallicity relation is quite flat for stars in the bulge region, indicating a wide range of stellar age populations at similar metallicities. To not find a difference in [M/H] distributions does not exclude significant age differences. However, according to \cite{Bensby2017}, the metal-poor and alpha-enhanced stars are still generally older than the metal-rich ones. From Figures \ref{fig_cnammh} (c1) and (c2), the [$\alpha$/M] and [M/H] distributions are very wide, suggesting the existence of multiple stellar populations in the bugle region ranging from young to old.

\subsection{Chemical abundances in Sgr fields}

In contrast to the bulge fields, the chemical distributions for the two components in the Sgr fields are significantly different. The HV peak is due to the core of the Sgr dSph, which lies $\sim$29 kpc from the Sun \citep{Siegel2011}, while the main component should be primarily composed of thick-disk stars along the line of sight toward the Sgr dSph. 
Note that the APOGEE selection function is different in the Sgr fields compared to the Galactic bulge region; the K magnitude of the Sgr field stars is systematically $\sim$1$-$2 mag fainter than stars in the bulge field. If a giant with the same luminosity is in the bulge (8.5 kpc) and the Sgr dSph  (29 kpc), the apparent magnitude difference is 2.5 mag. Toward the Sgr dSph, the extinction of Sgr stars is stronger than for bulge stars. Therefore, luminous giants are more likely included in the Sgr field, which may include more AGBs rather than RGBs. In this case, [C/N] may not be an accurate age proxy for AGBs in the Sgr HV peak.
Our metallicity distribution is consistent with that of   \cite{Cole2001} who suggested that the metallicity of stars in the Sgr dSph peaks at [Fe/H]$\sim -0.5 \pm 0.2 $ dex.
There are two scenarios to explain the observed chemical abundances: the mass-loss scenario and the starburst scenario \citep{McWilliam2005}. 
If the mass-loss scenario is dominant, a metallicity distribution biased to low-metallicity stars would be expected, while a discrete starburst might lead to a metallicity function dominated by a narrow peak. Our results seem to support the starburst scenario for the Sgr dSph, since an overdensity clump shows up in [M/H]$-$[C/N] and [M/H]$-$[$\alpha$/M] distributions.

The main component of the Sgr fields is largely composed of metal-poor stars. In the [M/H]$-$[C/N] distribution, it shows a different feature from the bulge stars.  Metal-rich stars of the Sgr main component spread over a narrower range compared to the bulge stars.  
The chemical enrichment history of Sgr is different from that of the MW. Separating Sgr from MW stars does not necessarily mean that one can separate stars of different ages from the MW bulge. However, for the MW bulge sample, if the metallicity distributions of the main component and the HV component cover a wide range from metal-poor to metal-rich stars, the stellar populations probably also span a wide age range. This is also supported by the recent results of \cite{Schiavon2016}, who used nitrogen and carbon abundances of APOGEE stars to disentangle a population of curious bulge field stars in the inner Galaxy, which are possibly dissolved globular cluster stars. \cite{Nidever2012} suggested that the HV peak stars in the bulge fields are not likely from the Sgr dSph because the velocity distribution is different from  their model prediction. We confirm this conclusion with our analysis of chemical abundance.

\subsection{Velocity distributions of different age populations}
\label{sec:section4.4}

\begin{figure*}[!t]

\begin{minipage}{0.48\linewidth}
\centerline{\includegraphics[width=1\textwidth]{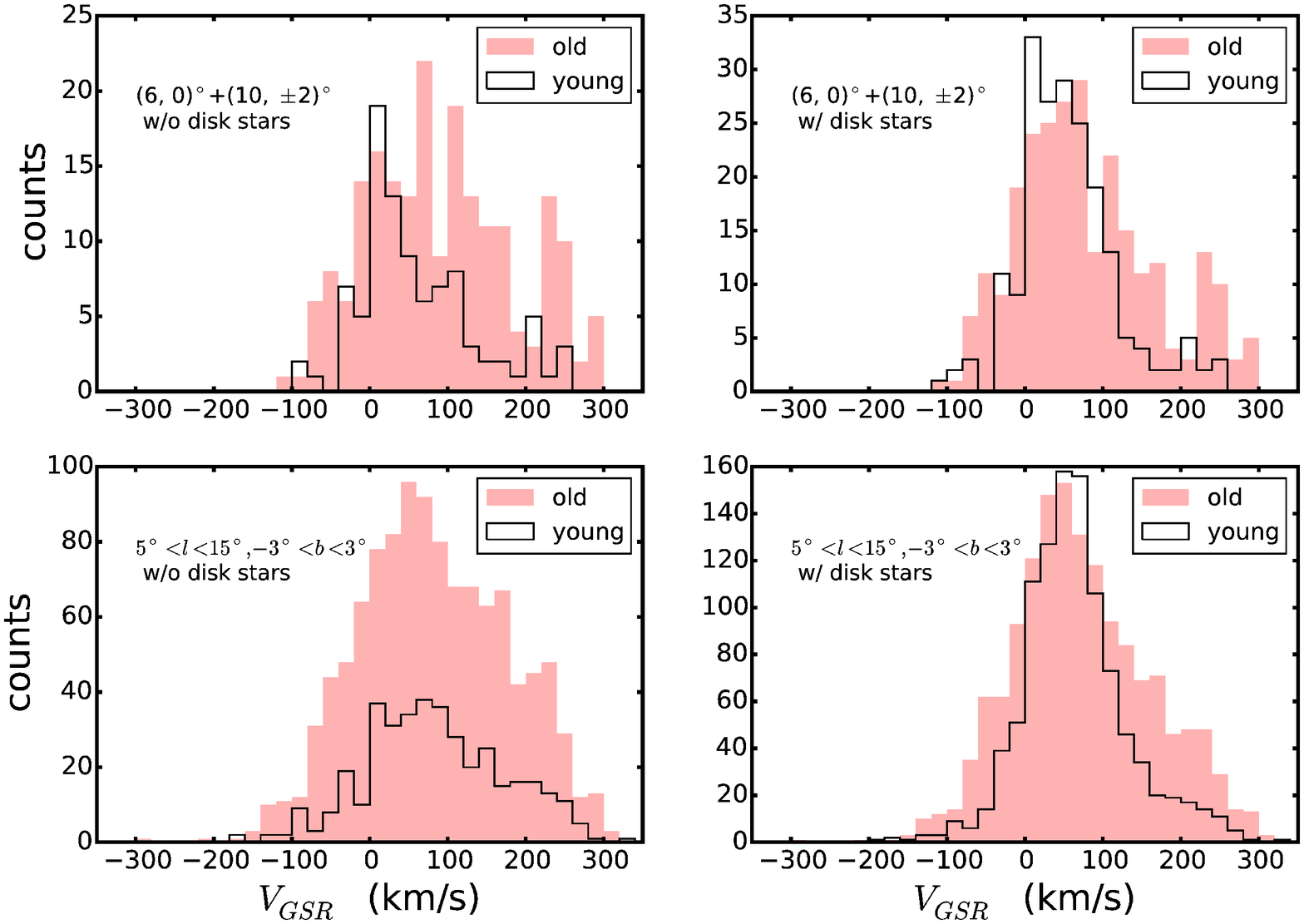}}
\caption{Velocity distributions of the young ([C/N]$<-$0.3, corresponding to roughly median age $<$3.5 Gyr; black histograms) and the old ([C/N]$>-$0.3, roughly median age $>$3.5 Gyr; red histograms) populations. The left columns show the results of our main sample (without foreground disk stars), and the right columns show the results including the foreground disk stars.}
\label{fig:vg_age_1}
\centerline{}
\end{minipage}
\qquad
\begin{minipage}{0.48\linewidth}
\centerline{\includegraphics[width=1\textwidth]{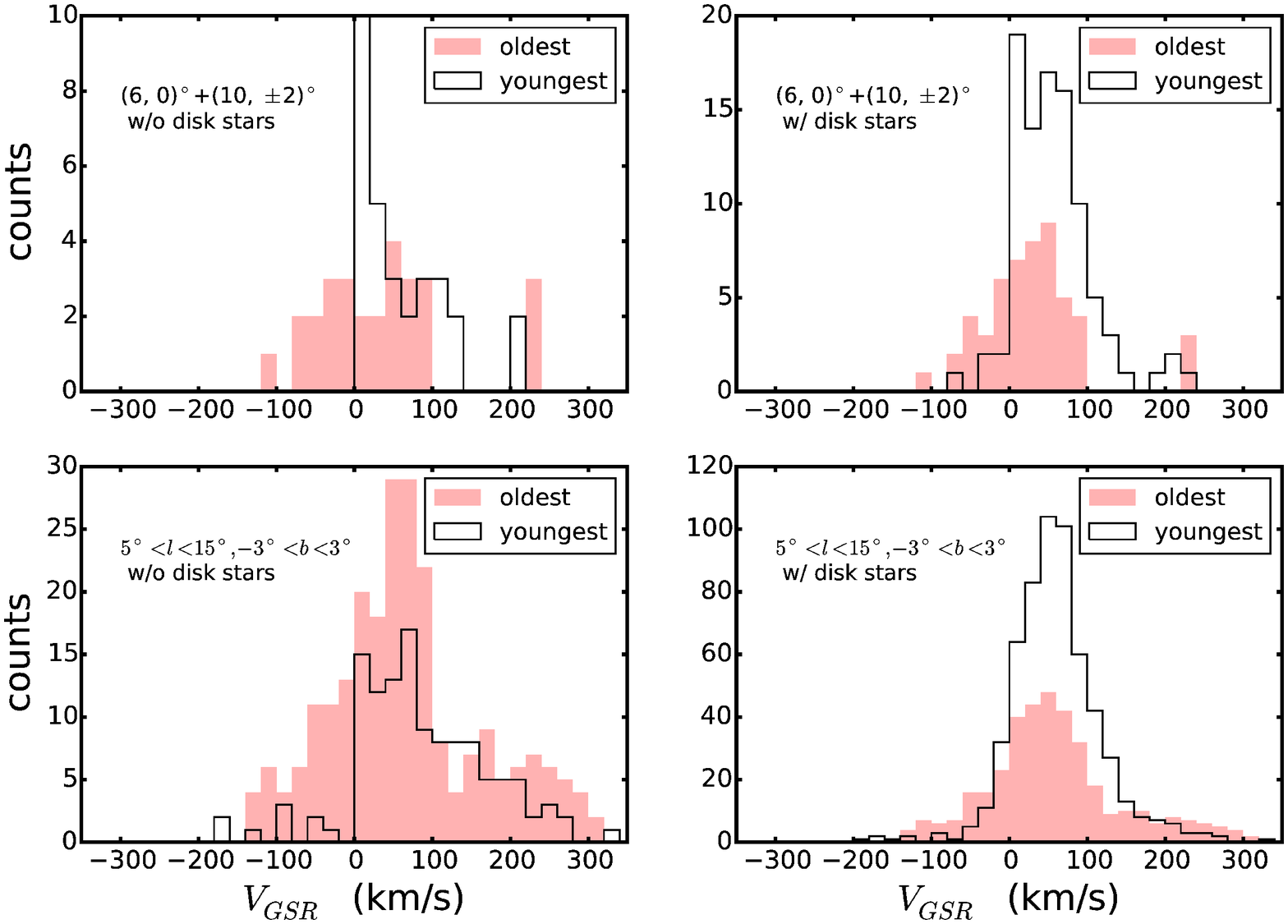}}
\caption{Velocity distributions of the  youngest ([C/N]$<-$0.4, median age $<$3 Gyr; black histograms) and the oldest ([C/N]$>-$0.1, median age $>$ 7 Gyr; red histograms) populations. The layout is the same as in Figure \ref{fig:vg_age_1}. }
\label{fig:vg_age_2}
\centerline{}
\end{minipage}
\end{figure*}

In their simulations \cite{Aumer2015} showed that only young bar stars show pronounced HV shoulders. Young stars might also contribute to produce a similar $h_{3}$ profile. To compare to these results directly,  we study the velocity distribution of the young and the old populations, using [C/N] as an age proxy. We consider the three fields showing clear HV peaks, i.e.  (6$^\circ$, 0$^\circ$), (10$^\circ$, $\pm$2$^\circ$), and the region showing high $h_{3}$ ($5^\circ<l<15^\circ$ and $-3^\circ<l<3^\circ$). The results are shown in Figure \ref{fig:vg_age_1}. 
Stars with [C/N] $<-$0.3 belong to the nominally young population (median age $<$3.5 Gyr), while stars with [C/N] $>-$0.3 belong to a nominally old population. In this test, the young stars generally have smaller velocity dispersion, as expected. However, the young population shows even weaker HV peak/shoulders than the old stellar population. The shape of the young stars' LOSVD seems roughly compatible with the \cite{Aumer2015} model, but the existence of a separate HV feature for the older stars is inconsistent with that model. We also test two extreme cases by selecting the youngest ([C/N] $<-$0.4, median age $<$3 Gyr) and the oldest ([C/N] $>-$0.1, median age $>$7 Gyr) populations in the sample (see Figure \ref{fig:vg_age_2}). From these figures we can see that the youngest population does not seem to make a large contribution to the HV peak, which is inconsistent with the suggestion of \cite{Aumer2015}.

\section{conclusion}

Cold, HV peaks in the bulge region were first reported by \cite{Nidever2012} using the commissioning data of APOGEE . We use the newly released APOGEE DR13 to revisit this result.
To understand better the bulge kinematics, for the first time foreground disk stars are excluded by applying a $T_{\rm eff}$ cut at 4000 K. In the LOSVDs, we find that most of the fields display a skewed Gaussian with an HV shoulder. However, only 3 out of 53 fields show a distinct HV peak. 

Most of the LOSVDs in the bulge fields can be well described by Gauss$-$Hermite series (except the three HV fields), and the spatial distribution of coefficient $h_{3}$ without removing foreground disk stars is largely consistent with the S10 model. We argue that the indicator of the bar is the positive correlation between $\bar{v}$ and $h_{3}$, which is consistent with predictions from bar models. This confirms that the Galactic bulge kinematics seen by APOGEE are dominated by a bar structure. Given the complicated bar kinematics, Gauss$-$Hemite moments can better describe the LOSVDs than fitting double Gaussian profiles. We also estimate the skewness of each LOSVD, and these show correlated profiles with the $h_{3}$ parameter. However, the skewness parameter can be significantly affected by the velocity outliers compared to $h_{3}$.

In the three bulge fields showing clear HV peaks, i.e., (6$^\circ$, 0$^\circ$) and (10$^\circ$, $\pm$2$^\circ$), the chemical abundances of the HV stars (including [M/H], [$\alpha$/M], and [C/N]) do not show any significant differences compared to the main component.   We further select stars in the region $5^{\circ} < l < 15^{\circ}$, $-3^{\circ} < b < 3^{\circ}$ to study the distributions of the main component and the HV component in [C/N]-[M/H] and [$\alpha$/M]-[M/H]. Again, the two components show similar chemical properties. Using [C/N] to trace age, we find that the age compositions of the two velocity components are similar and neither of them is dominated by young stars. This may be inconsistent with \cite{Aumer2015}, who suggested that the APOGEE selection function is  more sensitive to young stars and that the HV stars are dominated by a young population ($\lesssim$ 2 Gyr).

We separate the sample into a young and an old stellar population and find that the young population shows weaker HV peak/shoulders than the old stellar population.

We also find that the chemical abundance of Sgr core stars is different from that of bulge HV stars, so it is not likely that the HV peak stars are members of Sgr dSph. None of the models mentioned in the introduction (e.g., new halo structure, the tidal tail of Sgr, bar-supporting orbits, a kiloparsec-scale nuclear stellar disk and young stars captured by the bar) could explain the identified HV peaks well. The spatial distribution and chemical properties of the HV peak stars in the bulge region are not likely the tidal tail of Sgr or a new substructure in the halo. As shown in \cite{Lizhaoyu2014}, the full velocity distribution of the stars making up the bar potential can only produce a HV shoulder rather than a HV peak; the bar supporting orbits may not be able to produce a HV peak. A kiloparsec-scale nuclear disk proposed by \cite{Debattista2015} can explain the peak at $l = 8^{\circ} \pm 2^{\circ}$, but not off the midplane, which is inconsistent with observations. In this study, we show that the bulge stellar population is complicated with a wide range of age and metallicity. Moreover, both the young and old stellar populations show HV peak features, which are inconsistent with predictions in \cite{Aumer2015}. Other mechanisms are needed to explain the HV peaks, and the three observed peaks could be different in nature.

Forthcoming proper motions from Gaia observations of these bright bulge stars will help probe the orbits of the HV stars compared to "normal" bulge stars.

\acknowledgments

We thank the referee for the constructive and valuable comments that helped us to improve this paper. The research presented here is partially supported by the 973 Program of China under grant no. 2014CB845700; by the National Natural Science Foundation of China under grant nos.11333003, 11390372, 11322326, 11403072, 11373032, and by a China-Chile joint grant from CASSACA. J.S. acknowledges support from a {\it Newton Advanced Fellowship} awarded by the Royal Society and the Newton Fund and from the CAS/SAFEA International Partnership Program for Creative Research Teams. Z.-Y.L. is sponsored by Shanghai Yangfan Research
Grant (no. 14YF1407700). Z.-Y.L. acknowledges the support from the LAMOST Fellowship by Special Funding for Advanced Users, budgeted and administrated by the Center for Astronomical Mega-Science, Chinese Academy of Sciences (CAMS). This work made use of the facilities of the Center for High Performance Computing at Shanghai Astronomical Observatory. 
S.R.M. acknowledges support from NSF grants AST-1109718, AST-1312863, and AST-1616436. 
D.G. gratefully acknowledges support from the Chilean BASAL Centro de Excelencia en Astrof\'{i}sica
y Tecnolog\'{i}as Afines (CATA) grant PFB-06/2007. S.V. gratefully acknowledges the support provided by Fondecyt reg. no. 1170518. 

Funding for the Sloan Digital Sky Survey IV has been provided by the Alfred P. Sloan Foundation, the U.S. Department of Energy Office of Science, and the Participating Institutions. SDSS-IV acknowledges support and resources from the Center for High-Performance Computing at the University of Utah. The SDSS Web site is http://www.sdss.org.

SDSS-IV is managed by the Astrophysical Research Consortium for the Participating Institutions of the SDSS Collaboration, including the Brazilian Participation Group, the Carnegie Institution for Science, Carnegie Mellon University, the Chilean Participation Group, the French Participation Group, Harvard-Smithsonian Center for Astrophysics, Instituto de Astrof\'{i}sica de Canarias, The Johns Hopkins University, Kavli Institute for the Physics and Mathematics of the Universe (IPMU) / University of Tokyo, Lawrence Berkeley National Laboratory, Leibniz Institut f\"{u}r Astrophysik Potsdam (AIP), Max-Planck-Institut f\"{u}r Astronomie (MPIA Heidelberg), Max-Planck-Institut f\"{u}r Astrophysik (MPA Garching), Max-Planck-Institut f\"{u}r Extraterrestrische Physik (MPE), National Astronomical Observatories of China, New Mexico State University, New York University, University of Notre Dame, Observat\~{o}rio Nacional / MCTI, The Ohio State University, Pennsylvania State University, Shanghai Astronomical Observatory, United Kingdom Participation Group, Universidad Nacional Aut\'{o}noma de M\'{e}xico, University of Arizona, University of Colorado Boulder, University of Oxford, University of Portsmouth, University of Utah, University of Virginia, University of Washington, University of Wisconsin, Vanderbilt University, and Yale University.

\end{document}